\documentclass{elsart}
 \usepackage{epsfig}
\usepackage{amssymb}
\begin{document}

\newcommand{\EGLOB}{E_{\rm glob}}
\newcommand{\ELOC}{E_{\rm loc}}
\newcommand{\ERANDGLOB}{E_{\rm glob}^{\rm random}}
\newcommand{\ERANDLOC}{E_{\rm loc}^{\rm random}}
\newcommand{\be}{\begin{equation}}
\newcommand{\ee}{\end{equation}}
\newcommand{\bea}{\begin{eqnarray}}
\newcommand{\n}{\nonumber\\}
\newcommand{\eea}{\end{eqnarray}}

\begin{frontmatter}

\title{Is the Boston Subway a Small-World Network?}

\author[label1]
{Vito Latora}, 
\author[label2,label3]
{Massimo Marchiori}

\address[label1] {Dipartimento di Fisica e Astronomia,
Universit\'a di Catania,\\
and INFN sezione di Catania,Corso Italia 57 I 95129 Catania, Italy}

\address[label2] {W3C and Lab. for Comp. Science,
Massachusetts Institute of Technology, USA}

\address[label3]
{Dipartimento di Informatica, Universit\'a di Venezia, Italy}

\begin{abstract}
The mathematical study of the small-world concept has fostered 
quite some interest, showing that small-world features can be 
identified for some abstract classes of networks. 
However, passing to real complex systems, as for instance 
transportation networks, shows a number of new problems that
make current analysis impossible. In this paper we show how 
a more refined kind of analysis, relying on transportation
efficiency, can in fact be used to overcome such problems, 
and to give precious insights on the general characteristics 
of real transportation networks, eventually providing a picture 
where the small-world comes back as underlying construction principle.
\end{abstract}

\begin{keyword}
Small-World Networks \sep Transportation Systems
\end{keyword}
\end{frontmatter}

The characterization of the structural properties of the
underlying network is a very crucial issue to understand the
function of a complex system \cite{yaneer}. For example, the
structure of a social network affects spreading of information,
fashions, rumors but also of epidemics over the network; the
topological properties of a computer network (Internet, the World
Wide Web) affect the efficiency of the communication. Only
recently the accessibility of databases of real networks and the
availability of powerful computers have made possible a series of
empirical studies \cite{watts,bar1,stan1,lm1,stan2}. In
\cite{watts} Watts and Strogatz have shown that the {\it connection
topology} of some (social, biological and technological)
networks is neither completely regular nor completely random
\cite{watts}. Watts and Strogatz have named these networks, that
are somehow in between regular and random networks, {\it small
worlds}, in analogy with the small-world phenomenon observed in 
social systems \cite{milgram}. The mathematical characterization of the
small-world behavior is based on the evaluation of two quantities,
the characteristic path length $L$, measuring the typical
separation between two generic nodes in the network and the
clustering coefficient $C$, measuring the average cliquishness of
a node. Small-world networks are in fact highly clustered, like
regular lattices, yet having small characteristics path lengths,
like random graphs.

Although the initial small-world concept came from social networks, 
having a mathematical characterization makes it tempting to 
apply the same concept to any network representative of a 
complex system. This grand plan clashes with the fact that
the mathematical formalism of \cite{watts} suffers from severe 
limitations: 1) it applies only to some cases, whereas in general
the two quantities $L$ and $C$ are ill-defined; 2) it works only
in the {\it topological abstraction}, where the only information retained 
is about the existence or the absence of a link, and nothing is 
known about the physical length of the link.
\\
In this paper we take as paradigmatic example of real complex systems 
the realm of transportation (and use the Boston public
transportation system as real-world representative instance), 
showing how the passage from abstract 
social networks to applied complex systems present in nature poses
new challenges, that can in fact be overcome using a more general
formalism developed in ref.\ \cite{lm2} for {\it weighted networks}. 

\begin{figure}
\begin{center}
\epsfig{figure=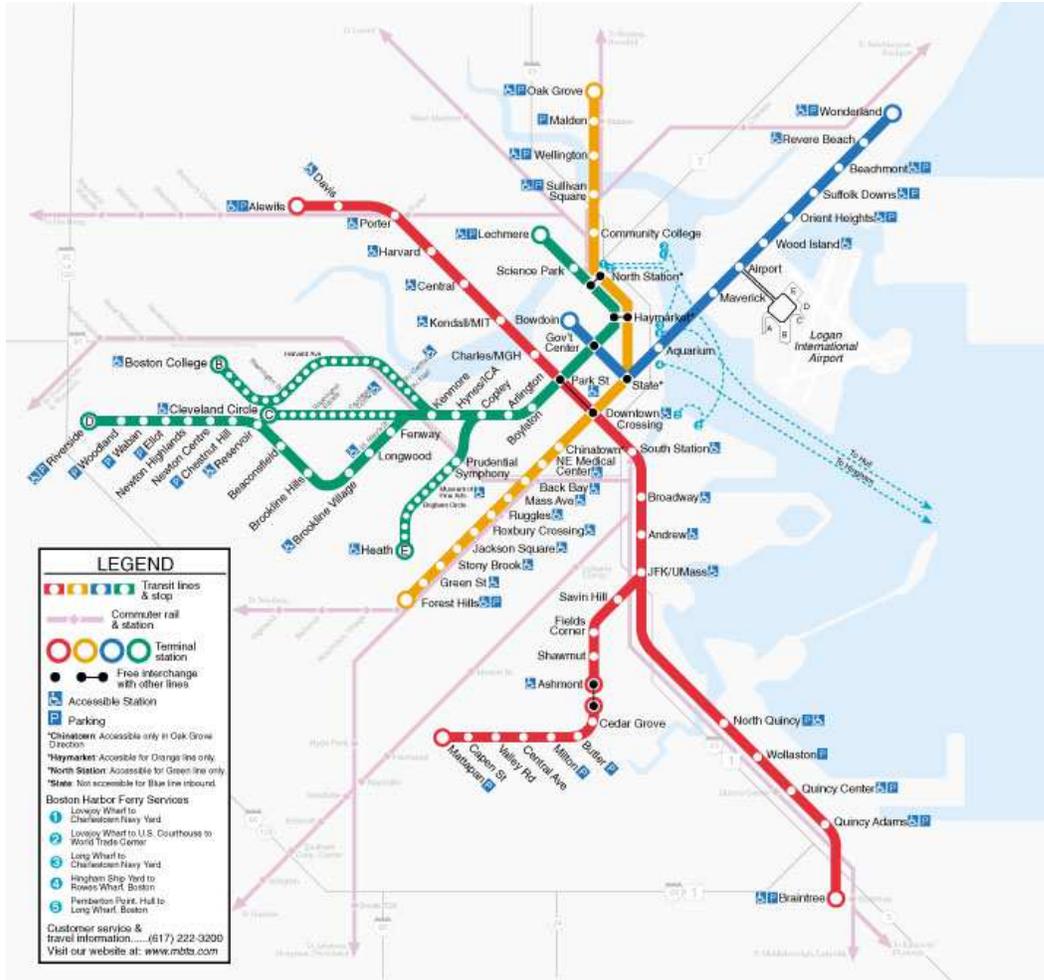,width=1.\columnwidth,angle=0}
\end{center}
\caption{The network of the {\em MBTA\/} consists of $N=124$
stations and $K=124$ tunnels. The matrix $\{\ell_{ij}\}$ has been
calculated using databases from the {\em MBTA\/}
\protect\cite{tboston} and the U.S.\ National Mapping Division. }
\end{figure}

The {\em MBTA\/} (Boston underground transportation system) 
consists of $N=124$ stations and $K=124$ tunnels
(connecting couples of stations) extending throughout Boston and
the other cities of the Massachusetts Bay \cite{tboston}. This
network can be considered as a graph with $N$ nodes and $K$ edges
and is represented by the adjacency (or connection) matrix $\{a_{ij}\}$, i.e.
the $N \cdot N$ matrix whose entry $a_{ij}$ is $1$ if there is an
edge joining node $i$ to node $j$ and $0$ otherwise, and by
$\{\ell_{ij}\}$ the matrix of the spatial (geographical) distances
between stations. According to the formalism of ref.\
\cite{watts}, valid for a subclass of unweighted (topological) networks, the
information contained in $\{\ell_{ij}\}$ is not used (as if
$\ell_{ij}=1 ~ \forall i \neq j$) and the shortest path length
$d_{ij}$ between two generic vertices $i$ and $j$ is extracted by
using only $\{a_{ij}\}$. The characteristic path length $L$ is the
average distance between two generic vertices:
$L= \frac {1}{N(N-1)} \sum_{i\neq j} d_{ij}$. 
The clustering coefficient $C$ is a local property defined as
follows. If the node $i$ has $k_i$ neighbors, then at most
$k_i(k_i-1)/2$ edges can exist between them; $C_i$ is the fraction
of these edges that actually exist, and $C$ is the average value
$C= \frac {1}{N} \sum_{i} C_i$. If we apply this method to try to study 
the MBTA, we obtain $L=15.55$ (an average of
15 steps, or 15 stations to connect 2 generic stations), while $C$
is not well defined since there are few nodes nodes with only 1
neighbours, and then $C_i =\frac{0}{0}$ for these nodes. 
In any case to decide if the MBTA is a
small world we have to compare the $L$ and $C$ obtained to the
respective values for a random graph with the same $N$ and $K$.
When we consider a random graph we incur into
the same problem for $C$; moreover we get $L=\infty$ because in
most of the realizations of the random graph there are some nodes
not connected to the remaining part of the network. Summing up, by 
mean of $L$ and $C$ we are unable to draw any conclusion.
\\
Now we propose our alternative formalism (based on ref.\ \cite{lm2}), 
valid for weighted, and also disconnected networks. The
matrix of the shortest path lengths $\{d_{i,j}\}$ is now
calculated by using the information contained both in $\{a_{ij}\}$
and in $\{l_{ij}\}$.
Instead of $L$ and $C$, the network is characterized in
terms of how efficiently it propagates information on a global and
on a local scale respectively. 
We assume that the efficiency $\epsilon_{ij}$ in the communication
between node $i$ and $j$ is inversely proportional to the shortest
distance:  $\epsilon_{ij} = 1/d_{ij} ~\forall i,j$. We see
immediately that this way we avoid the problem of the
divergence we had for $L$, in fact when there is no path in the
graph between $i$ and $j$, $d_{i,j}=+\infty$ and consistently
$\epsilon_{ij}=0$. Moreover, the link characteristics 
(length/capacity in the case of transportation systems) are properly
taken into account, and not flattened into their topological abstraction. 
We define the network {\it efficiency} as $ E =
\frac{1}{N(N-1)} \sum_{i \ne j} \epsilon_{ij}
           =  \frac{1}{N(N-1)} \sum_{i \ne j} \frac{1}{d_{ij}} $.
The quantity E is normalized to the efficiency of the ideal case
in which the network has all the $N(N-1)/2$ possible edges:
in this way $0 \le E \le 1$ \cite{lm2}.
We call $\EGLOB$ the efficiency of the whole network and
$\ELOC$ the average efficiency of the subgraph
of the neighbors of a generic node $i$.
In ref.\cite{lm2} we have shown that $\EGLOB$ and $\ELOC$ play
respectively the role of $L$ and $C$, and that small-world
networks have both high $\EGLOB$ and  high $\ELOC$. 
\\
Now, let us apply these new measures to the MBTA: the results 
are reported in tab.1.
As we can see, the {\em MBTA\/} turns out to be a very efficient 
transportation system on a global scale but not at the local level. 
Let us analyze better what insights the calculation shows. 
In fact, $\EGLOB=0.63$ means that {\em MBTA\/} is only $37\%$
less efficient than the ideal subway with a direct tunnel
from each station to the others, quite a remarkable result.
\begin{table}
\caption{~~~ Global and local efficiency and cost of the MBTA. In
the first row the MBTA is considered. In the second 
raw the composite system MBTA+bus is considered. \label{table1}}

\begin{tabular}{l|ll|ll|l}
                    & $\EGLOB$   & $\ELOC$ & $Cost~~$\\
MBTA     & 0.63       & 0.03    & 0.002   \\
MBTA+bus & 0.72       & 0.46    & 0.004
\end{tabular}
\end{table}
On the other hand, $\ELOC=0.03$ indicates a poor local efficiency: 
this shows that, differently from social systems, the {\em MBTA\/} 
is not fault tolerant and a damage in a station
will dramatically affect the efficiency in the connection between
the previous and the next station.
In order to better understand the difference with respect to other 
systems that are globally but also 
locally efficient we need to consider the cost of a network.
In general we expect the efficiency of a network to
be higher when the number of edges increase.
As a counterpart, in any real network
there is a price to pay for number and length (weight) of edges.
To quantify this effect we define the {\it cost} of a network
as: $Cost = {\sum_{{i \ne j}} a_{ij} \ell_{ij}}
/ {\sum_{{i \ne j}} {\ell_{ij}}}$. We have $0 \le Cost \le 1$, and the
the maximum value $1$ is obtained for the ideal case
when all the edges are present in the network.
$Cost$ reduces to the normalized number of edges $2K/N(N-1)$
in the case of an unweighted graph.
For the {\em MBTA\/} we get an extremely small value $Cost=0.002$. 
This means that {\em MBTA\/} achieves the $63\%$ of the efficiency
of the ideal subway with a cost of only the $0.2\%$.
Qualitatively similar results have been obtained for
other underground systems. 
The price to pay for such low-cost high
global efficiency is the lack of fault tolerance.
This means that when we build a subway system,
the priority is given to the achievement of global efficiency at a
relatively low cost, and not to fault tolerance. 
But where is the rationale for such a construction principle? 
In fact, fault tolerance in such a transportation system is less of a
critical issue as it would seem: a temporary problem in a station can
be solved in an economic way by other means, for example by taking a bus
from the previous to the next station.
That is to say, lack of fault tolerance for users is only apparent: 
the MBTA is not a {\em closed system\/}, as 
it can be considered, after all, a subgraph of a wider
transportation network, and this explains why,
fault tolerance is not a critical issue.
Changing the MBTA network to take into account for example the bus
systems, indeed, shows that this extended transportation system 
is a small-world network ($\EGLOB=0.72$, $\ELOC=0.46$)!
Therefore, efficiency and fault-tolerance come back as a leading 
underlying construction principle, and the whole transportation
system MBTA+bus turns out to be a small-world with a 
slight increase in the cost ($Cost=0.004$).

Summing up, the analysis of real-life complex systems like transportation 
networks poses a number of new challenges, that make the initial 
mathematical formalization of small worlds in ref.\ \cite{watts} fail. 
The introduction of the efficiency measure 
allows to give a more general mathematical definition of small worlds, able
to deal successfully with transportation systems (and in general, 
for weighted networks). Such measure, like in the MBTA case, provides
quantitative information on the efficiency characteristics of a system,
helping to explain the underlying construction principles. 
Moreover, apparent lack of a generalized small-world behaviour can, 
as in the MBTA case, be explained by the fact we have just a partial
view of the complete system. In fact, the analysis presented
in this paper shows that a generic {\em closed\/} transportation system can 
exhibit the small-world behavior, substantiating the idea that, in 
the grand picture, the diffusion of small-world networks can be interpreted 
as the need to create networks that are both globally and locally efficient.

\bigskip
\noindent

\end{document}